\documentclass[reprint,twocolumn,showpacs,amsmath,amssymb,superscriptaddress,prl]{revtex4-1}

\usepackage[dvips]{graphicx}
\usepackage{dcolumn}
\usepackage{bm}
\usepackage{amsmath,amsthm,amssymb}

\newcommand{\affA}{%
    Van der Waals-Zeeman Institute, University of Amsterdam,\\
Valckenierstraat 65, 1018 XE Amsterdam, The Netherlands }

\begin{document}

\title{Sub-Poissonian atom number fluctuations by three-body loss in mesoscopic ensembles}

\author{S. Whitlock}
\author{C.~F. Ockeloen}
\author{R.~J.~C. Spreeuw}\affiliation{\affA}
\email{S.M.Whitlock@uva.nl}

\date{\today}

\begin{abstract}
We show that three-body loss of trapped atoms leads to sub-Poissonian atom number fluctuations. We prepare hundreds of  dense ultracold ensembles in an array of magnetic microtraps which undergo rapid three-body decay. The shot-to-shot fluctuations of the number of atoms per trap are sub-Poissonian, for ensembles comprising 50--300 atoms.  The measured relative variance or Fano factor $F=0.53\pm 0.22$ agrees very well with the prediction by an analytic theory ($F=3/5$) and numerical calculations. These results will facilitate studies of quantum information science with mesoscopic ensembles.
\end{abstract}

\pacs{03.75.Be, 05.40.-a, 42.50.Lc, 67.85.-d} 

\keywords{}

\maketitle

The study and control of particle number fluctuations in ultracold atomic systems has revealed a rich variety of intriguing quantum phenomena~\cite{GreRegJin05,FolGerBlo05,RomBesBlo06,GerFloBlo06,JelMcNWes07,GemZhaChi09}, and offers the potential to boost performance in cold atom technologies. Motivated largely by the prospects for quantum metrology~\cite{WinBolHei94,SorDuaZol01,GioLloMac04}, several recent experiments have demonstrated the suppression of \emph{relative} fluctuations between small atomic samples distributed over two or more traps or internal states, leading to number difference or spin squeezing and entanglement~\cite{JoShiPre07,EstGroObe08,AppWinPol09,SchLerVul0810.2582}. By contrast, however, work on suppressing \emph{absolute} number fluctuations has been limited~\cite{ChuSchRai05,ItaVekSte0903.3282}. This is crucial, for example, in quantum information science using mesoscopic atomic ensembles~\cite{LukFleZol01,BriMolSaf07,SafMol09,MulLesZol09}, where recently observed collective excitations produced via Rydberg dipole blockade~\cite{HeiRaiPfa07,UrbJohSaf09,GaeMirGra09} could be exploited. Trapped ensembles would benefit from a $\sqrt{N}$ collective enhancement of the Rabi frequency over single atoms, allowing fast quantum operations. However, intrinsic atom number fluctuations would adversely affect the fidelity. 

In this Letter we show explicitly that three-body loss naturally reduces the shot-to-shot fluctuations of the absolute atom number in a trap to sub-Poissonian levels. In experiments, random particle loss is usually considered deleterious, and it is not generally recognized that random loss can suppress fluctuations, even below the Poisson level. This is the atomic analog to intensity squeezing in optics~\cite{Man82,Lou84,Hil87,GilKni93}. We show that three-body loss can be used to prepare small and well-defined numbers of atoms in each trap, ultimately enabling the study of collective excitations in mesoscopic ensembles. We trap a large number of dense mesoscopic ensembles in a lattice of microtraps which undergo rapid three-body decay. Through sensitive absorption imaging we measure the shot-to-shot distribution of atom numbers and find sub-Poissonian statistics for between $50$ and $300$ atoms per trap. The effects of residual imaging noise are greatly reduced through the application of spatial correlation analysis which exploits the lattice geometry and provides a way to isolate atom number fluctuations. Our results are in very good agreement with a model for stochastic three-body loss which takes into account the fluctuations. 


For ultracold gases in magnetic microtraps, inelastic density-dependent decay is the dominant loss process. In $^{87}$Rb this is typically due to three-body recombination~\cite{MoeBoeVer96,EsrGreBur99,NorBalBra06}, whereby all three atoms are lost from the trap. As this depends on the probability of finding three atoms together, three-body recombination is a sensitive probe of density fluctuations and correlations in degenerate Bose-gases~\cite{BurGhrWie97,SodGueDal99,TolOHaHuc04}. This previous work involved the macroscopic evolution of the {\emph{mean number}} of remaining atoms, which decays proportional to the mean square density. 

We are primarily interested in the {\emph{fluctuations}} in the number of remaining atoms. We model this with the following master equation for the probability distribution $P(N,t)$,
\begin{eqnarray}
    \frac{dP(N,t)}{dt} =
      \sum_{\rho=1,2,3}
      \frac{k_\rho\bigl{(}\mathbb{E}^\rho-1\bigr{)}}{\rho N_0^{\rho-1}}
      \frac{N!}{(N-\rho)!}P(N,t),
\label{eq:meqn}
\end{eqnarray}
which is valid for any birth-death process with multiple reactions involving $\rho$ bodies \cite{B:vanKampen07}. Here $N_0$ is the initial mean atom number in a given trap, $k_\rho$ are the scaled rate constants and the step operator $\mathbb{E}^\rho$ changes $N\rightarrow N+\rho$. Eq.~\eqref{eq:meqn} is a set of coupled differential equations, one for each possible value of $N$. For small systems involving up to a few hundred atoms, these equations can be solved numerically to provide the full atom statistics (including fluctuations) as a function of time.

In our experiments $k_2\approx\nobreak 0$ and $k_3\gg k_1$. For a non-degenerate gas at temperature $T$ in a harmonic trap, $k_3/N_0^2=\nobreak (2L_3/\sqrt{3})(m\bar{\omega}^2/2\pi k_B T)^{3}$, where the mean trap frequency in our case is $\bar{\omega}=\nobreak 2\pi\times~10.0\pm\nobreak 0.5$~kHz. The three-body rate constant is $L_3= 1.8(\pm 0.5)\times10^{-29}$~cm$^6$/s for the $\mathcal{F}=m_\mathcal{F}=2$ hyperfine state of $^{87}$Rb~\cite{SodGueDal99}.

For the mean and variance of the distribution we can obtain approximate analytic expressions. Following~\cite{B:vanKampen07} we perform a system size expansion for $N_0\gg 1$ to obtain a linear Fokker-Planck equation and derive equations of motion for the moments. For combined one-body and three-body loss the evolution of the mean fraction of remaining atoms is
\begin{eqnarray}
    \eta=\frac{\langle N\rangle}{N_0}=\frac{\exp({-k_1 t})}{\sqrt{1+(k_3/k_1)\left[1-\exp({-2 k_1 t})\right]}}.
\label{eq:meanNdecay}
\end{eqnarray}

We express the fluctuations in terms of the relative variance, or Fano factor, $F=(\langle N^2\rangle-\langle N\rangle^2)/\langle N\rangle$, where the averages are taken over realizations ($F=1$ for a Poisson distribution). The evolution of $F$ in time can be written as a function of $\eta$. This leads to the following differential equation:
\begin{equation}
    \frac{dF}{d\eta}=\frac{k_3 \eta^2(5F(\eta)-3)+k_1(F(\eta)-1)}{\eta(k_1+k_3\eta^2)}.
\label{eq:dFdN}
\end{equation}
In the case where three-body loss dominates we obtain the simple solution
\begin{equation}
    F(\eta)=\frac{3}{5}+\eta^5\left(F_0-\frac{3}{5}\right),
\end{equation}
where $F_0=F(\eta=1)$ is the initial Fano factor. As the atoms are lost from the trap the Fano factor asymptotes to a value of $F\rightarrow3/5$, significantly below the Poissonian level $F=\nobreak 1$. Correspondingly the memory of the initial Fano factor is lost very rapidly due to the fifth power of $\eta$, in contrast to one-body loss where $F=1+\eta (F_0-1)$. The result is easily generalized to an arbitrary $\rho$-\nobreak body process yielding an asymptotic Fano factor $F\rightarrow\nobreak \rho/(2\rho-1)$. The results of this simple analytic model are in excellent agreement with the numerical solution to Eq.~\eqref{eq:meqn} for $\langle N\rangle\gtrsim10$.

\begin{figure}%
\includegraphics[width=\columnwidth]{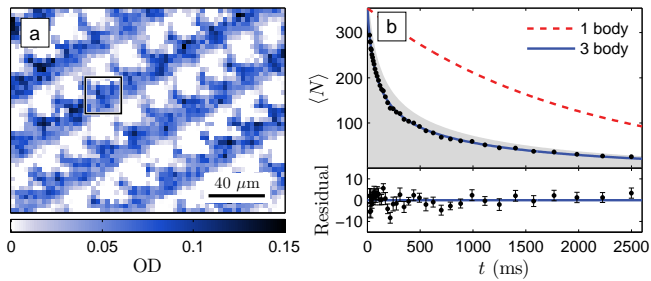}
\caption{(a) Subsection of an optical density image for $t=25$~ms for a single realization of the experiment. (b) Decay of the mean atom number in a selected site (highlighted in (a)). A fit to Eq.~\eqref{eq:meanNdecay} is shown (solid line) together with the corresponding one-body loss (dashed line). The shaded region indicates the range of atom numbers in our data for all traps. The residuals from the fit and standard errors on the measurement of $\langle N\rangle$ are shown below, demonstrating agreement at the level of $\pm 3$ atoms. }
\label{fig:decay}%
\end{figure}

Our experiment incorporates a two-dimensional lattice of optically resolvable magnetic microtraps produced by a magnetic film atom chip~\cite{GerWhiSpr07,WhiGerSpr09}. We load a few thousand atoms into each of approximately 250 traps, and then evaporatively cool close to quantum degeneracy (temperature $\sim3~\mu$K, phase-space-density $\sim0.3$). After cooling, a few hundred atoms remain in each trap. Due to the small size of each trap the atomic density is high ($\approx 2\times 10^{14}$~cm$^{-3}$) and we observe rapid three-body loss, despite relatively few atoms per site. During the experiment we apply a fixed radio-frequency `knife' (effective trap depth $\sim35~\mu$K) to ensure the temperature of each cloud does not vary. The knife counteracts any heating that may accompany the three-body loss and due to the high trap depth the role of heating-induced loss on the expected fluctuations is negligible.

We image the in-situ distribution of atoms (Fig.~\ref{fig:decay}a) using absorption imaging in reflection geometry with a circularly polarized probe laser aligned perpendicular to the chip surface~\cite{WhiGerSpr09}. The effective pixel size in the object plane is $3.2~\mu$m, the optical resolution is $7.5~\mu$m (Rayleigh criterion) and the lattice spacings are $22~\mu$m and $36~\mu$m. The exposure time is 0.15~ms and the saturation parameter is $s=2\times0.3$ (double pass). In each run of the experiment we record an absorption image, a reference image taken without atoms and a stray light image, from which we compute an optical density image of the atomic distribution. Each image contains the center-most region of the loaded lattice and a surrounding background region used to quantify the imaging noise.

To measure the decay we hold the atoms for a variable time after evaporative cooling before taking the absorption image. Our data are comprised of two sets. The first spans from $t=0~$ms to $t=880$~ms with 40 intervals (selected on a power-law scale) and repeated 15 times (600 runs of the experiment). The second data set spans from $t=21$~ms to $t=2.5$~s, with 40 intervals and 19 repeats (760 runs). From this we extract for each microtrap (with index $m$) (i) the decay of the mean atom number $\langle N_m\rangle$ and (ii) the variance $\langle N_m^2\rangle-\langle N_m\rangle^2$. Each optical density image is aligned to the average to minimize the effect of jitter between shots. The atom number in each site and each image is found by a two-dimensional amplitude fit of a model shape function which minimizes the influence of imaging noise. The shape functions are obtained by Gaussian decomposition of the average optical density image, for each cell of the lattice. The obtained shapes are smooth peaked functions which reproduce the observed absorption profiles (Fig~\ref{fig:decay}(a)), accounting for small distortions due to the underlying chip surface. Least-squares amplitude fitting is then performed on each image for 245 ensembles, with each fit including the 8 nearest neighbors to account for small overlapping areas.

\begin{figure}[]
\centering\includegraphics[width=0.95\columnwidth]{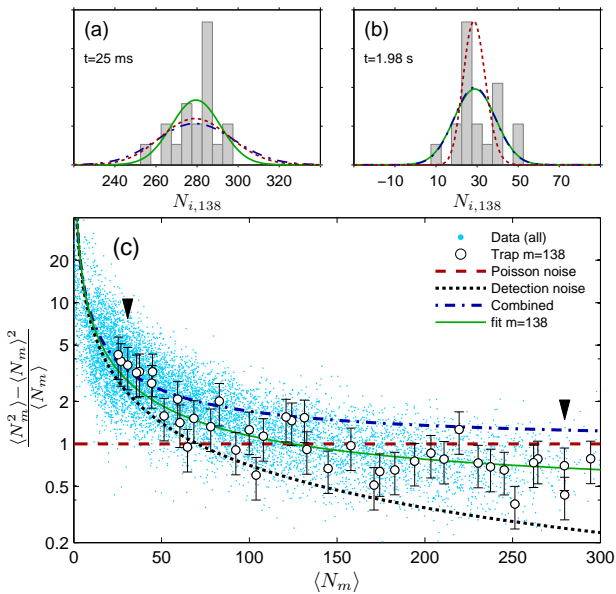}
\caption[Atom number fluctuations measured for each of 245 lattice sites during three-body decay.]{Atom number fluctuations measured for each of 245 lattice sites during three-body decay. (a,b) Number distributions for one specific trap ($m$=138) at two hold times. Histograms correspond to 19 measurements and each bin is $5~$atoms wide. The lines indicate Gaussian fits to the data (solid), Poisson distributions (dashed), and combined Poisson and detection noise contributions (dash-dotted). (c) The relative variance vs. $\langle N_m\rangle$ for each lattice site and for each hold time (points). Open circles indicate the measurements for the selected trap, with a fit (including detection noise) for a constant Fano factor $F=0.57$ (solid line). Arrows highlight the two data points corresponding to the histograms (a,b).}
\label{fig:shot-to-shot}
\end{figure}

Figure~\ref{fig:decay}a shows a section of an optical density image for a hold time of $25~$ms. The evolution of $\langle N\rangle$ for a selected trap is shown in Fig.~\ref{fig:decay}b. A fit to the data with Eq.~\eqref{eq:meanNdecay} yielding $k_3=10.4\pm0.4~$s$^{-1}$, $k_1=0.52\pm0.03~$s$^{-1}$ and $N_0=354\pm4$ is shown, together with the corresponding one-body decay $\eta = \exp(-k_1t)$. The shaded region indicates the full range of atom numbers in our data obtained by analyzing each trap individually.

To quantify the fluctuations it is necessary to accurately calibrate the absorption cross-section. For this we compare for each trap individually the measured cloud temperature and three-body loss rate~\cite{WhiGerSpr09}, to independently infer the atom number. In this way we determine an absorption cross section of $(0.32\pm0.05)\sigma_0$ ($\sigma_0=3\lambda^2/2\pi$) which is in good agreement with the expected cross-section of $0.31\sigma_0$ based on our imaging parameters. The maximum optical depth for a trap containing $250$ atoms is $\sim0.1$. 



Figure~\ref{fig:shot-to-shot} shows the measured atom number statistics for various hold times, corresponding to different mean atom numbers in each trap. A histogram of the fitted number of atoms in one specific trap for 19 repetitions of the experiment at $t=25$~ms is shown in Fig.~\ref{fig:shot-to-shot}a. The measured $\langle N\rangle=280\pm3$ and the variance is $\langle N^2\rangle-\langle N\rangle^2=140\pm50$, indicated by the Gaussian distribution (solid line). The distribution is significantly narrower than for a Poisson distribution (dashed line), providing a direct observation of sub-Poissonian number statistics in our experiment. For longer hold times (Fig.~\ref{fig:shot-to-shot}b) the mean number of atoms decreases due to loss, however the observed distribution does not become significantly narrower. This is due to the added detection noise contribution (dash-dotted line) which begins to dominate the observed fluctuations for $\langle N\rangle\lesssim 60$.

The same analysis is performed for each site and each hold time independently to obtain the site-resolved relative variance as a function of the mean number of atoms. Fig.~\ref{fig:shot-to-shot}c shows the results of $245\times40$ observations where each point is derived from 19 measurements. The observed fluctuations have two main contributions, atom noise with a constant $F$ (Poisson noise is indicated by a dashed line) and a detection noise contribution corresponding to a fixed variance of 64~atoms$^2$/trap/shot (dotted line). We find for $N\gtrsim100$ the vast majority of data points fall well below the combined variance for Poisson fluctuations (dash-dotted line) indicating $F<1$. Interestingly the deviation from Poisson statistics is most apparent for small hold times (large $\langle N\rangle$), indicating three-body loss also has a significant effect on the fluctuations before the end of the evaporative cooling stage.


\begin{figure}[]
\centering\includegraphics[width=0.95\columnwidth]{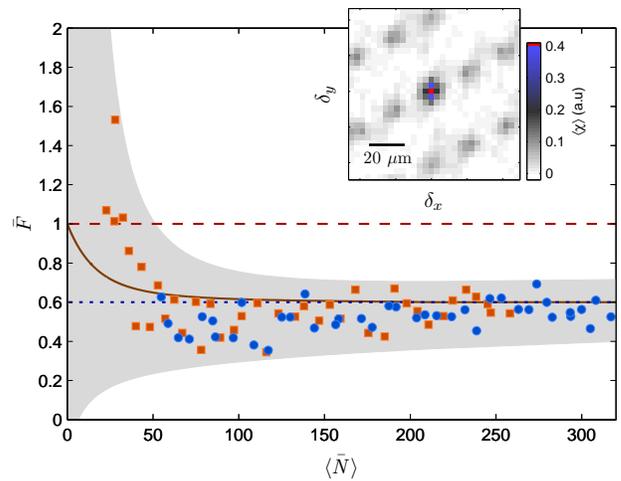}
\caption[Lattice averaged Fano factor $\bar{F}$ as a function of the mean number of atoms per site]{Lattice averaged Fano factor $\bar{F}$ as a function of the mean number of atoms $\langle\bar{N}\rangle$. Horizontal lines correspond to $F=1$ (dashed) and to $F=3/5$ (dotted) for strong three-body loss. The solid line is a model including three-body and one-body loss terms. The shaded region indicates systematic uncertainties described in the text. The inset shows an example fluctuation correlation function $\langle \chi(\bm{\delta})\rangle$ for $t=25$~ms.}
\label{fig:fcorranalysis}

\end{figure}
To account for detection noise and to investigate the sub-Poissonian noise over the full range of atom numbers in our experiment we perform spatial correlation analysis of the optical density images. Here we benefit from the lattice geometry and separate various noise components based on their respective correlation length-scales to isolate the atom fluctuations in our data.

We compute, for each optical density image the two-dimensional fluctuation correlation function $\chi_i({\bm{\delta}})\!=\!\nobreak\int\nobreak (n_i({\bm{x}})\!-\!\langle n_i({\bm{x}})\rangle)(n_i({\bm{x}}\!+\!{\bm{\delta}})\!-\!\langle n_i({\bm{x}}\!+\!{\bm{\delta}})\rangle) d^2{\bm{x}}$, which is then averaged over the realizations of the experiment (indexed by $i$) for a given hold time (Fig.~\ref{fig:fcorranalysis}~inset). We model the observed spatial distribution by $n_i(\bm{x})=c_i\sum_m N_{i,m}p_m(\bm{x})+d_i(\bm{x})$, where $N_{i,m}$ and $p_m(\bm{x})$ are the number of atoms and local shape function respectively for ensemble $m$, $c_i\approx 1$ accounts for correlated noise (due for example to probe frequency noise) and $d_i(\bm{x})$ accounts for spatially uncorrelated imaging noise. The correlation function $\langle\chi(\bm{\delta})\rangle$ shows several distinct features (Fig.~\ref{fig:fcorranalysis}~inset). A narrow spike at $\bm{\delta}=0$ (central red pixel) represents the uncorrelated imaging noise. This sits on top of a broader peak (dark central feature) representing the fluctuations correlated over the length scale of approximately a single cloud which accounts for shot-to-shot fluctuations of the number of atoms within each trap. An array of neighboring peaks, spaced at the lattice period, represents the correlated noise across traps which we attribute to small fluctuations of the probe detuning.

In the analysis of $\langle\chi(\bm{\delta})\rangle$ for each hold time, we first subtract the calculated background-region correlation function and exclude the $\bm{\delta}=0$ pixel spike to eliminate the uncorrelated imaging noise. We then fit two-dimensional Gaussian distributions to the central and neighboring correlation function peaks. The lattice-averaged Fano factor (weighted by $\langle N_m\rangle$) is given by $\bar{F}=(\sum_m\langle N_m^2\rangle-\nobreak\sum_m\langle N_m\rangle^2)/\sum_m\langle N_m\rangle$. Neglecting the small overlap between neighboring shape functions and noting that the fluctuations of $N_{i,m}$ are uncorrelated between different traps, we obtain
\begin{equation}
    \bar{F}=\left(\frac{X_0-X_\lambda}{X_\lambda+P_0}\right) \,\langle\bar{N}\rangle,
\end{equation}
where $X_0$ and $X_\lambda$ are the fitted volumes of the central and neighboring peaks of $\langle\chi({\bm{\delta}})\rangle$ respectively, $P_0$ is the fitted volume of the pre-averaged autocorrelation function peak $\int\nobreak \langle n_i({\bm{x}})\rangle\langle n_i({\bm{x}}+{\bm{\delta}})\rangle d^2{\bm{x}}$, and $\langle\bar{N}\rangle$ is the weighted average atom number. Accounting for the overlap between neighboring shape functions yields a small correction factor, which for our lattice geometry is $\lesssim 1.1$.

Figure \ref{fig:fcorranalysis} shows the extracted Fano factor for two separately analyzed data sets as a function of $\langle\bar{N}\rangle$ during the hold time. Horizontal lines correspond to the Poissonian limit $\bar{F}=1$ (dashed) and to the expected limit $\bar{F}=3/5$ (dotted) for strong three-body decay. The data shows sub-Poissonian atom number fluctuations for $\langle\bar{N}\rangle\geq50$ up to 300 atoms per site. A fit over this range indicates a Fano factor of $\bar{F}=0.53$ with a standard deviation of $\pm 0.08$. We independently estimate a systematic uncertainty of $\pm 0.2$ incorporating uncertainties in the absorption cross-section, background noise contribution and the overlap between neighboring traps. The measured fluctuations are clearly below the Poissonian noise level (dashed-line) and are in good agreement with the theoretical expectation of $\bar{F}=3/5$ (dotted-line). For $\langle\bar{N}\rangle\leq 50$ one-body loss dominates and we expect $\bar{F}$ to increase to 1. The solid line is the result of Eq.~\eqref{eq:dFdN} including both three-body and one-body loss terms.


In conclusion, we have shown that normally-undesirable density dependent losses in small atomic ensembles naturally lead to suppressed fluctuations of the absolute atom number to below Poissonian noise levels. By three-body decay it is possible to prepare hundreds of small and well-defined atomic ensembles consisting of tens to a few hundred atoms. We expect this to be an ideal system for the study of collective excitations produced for example via laser-excited Rydberg states for quantum information processing with neutral atoms~\cite{LukFleZol01,BriMolSaf07,MulLesZol09,SafMol09}. Such ensembles also have desirable properties for generation of Schrodinger-cat-like states~\cite{MulLesZol09}, the study of spin-squeezing and as a resource for quantum metrology using trapped atoms~\cite{AppWinPol09,SchLerVul0810.2582}. 
\begin{acknowledgments}
We would like to thank N. J. van Druten and J. T. M. Walraven for fruitful discussions. We are grateful to FOM and NWO for financial support. 
SW acknowledges support from a Marie-Curie fellowship (PIIF-GA-2008-220794).
\end{acknowledgments}
\bibliography{amo2,books}

\begin{thebibliography}{10}

\bibitem{GreRegJin05}
M. Greiner, C.~A. Regal, J.~T. Stewart, and D.~S. Jin, Phys. Rev. Lett. {\bf
  94},  110401  (2005).

\bibitem{FolGerBlo05}
S. F\"{o}lling {\it et~al.}, Nature {\bf 434},  481  (2005).

\bibitem{RomBesBlo06}
T. Rom {\it et~al.}, Nature {\bf 444},  733  (2006).

\bibitem{GerFloBlo06}
F. Gerbier {\it et~al.}, Phys. Rev. Lett. {\bf 96},  090401  (2006).

\bibitem{JelMcNWes07}
T. Jeltes {\it et~al.}, Nature {\bf 445},  402  (2007).

\bibitem{GemZhaChi09}
N. Gemelke, X. Zhang, C.-L. Hung, and C. Chin, Nature {\bf 460},  995  (2009).

\bibitem{WinBolHei94}
D.~J. Wineland, J.~J. Bollinger, W.~M. Itano, and D.~J. Heinzen, Phys.\ Rev.~A
  {\bf 50},  67  (1994).

\bibitem{SorDuaZol01}
A. {S{\o}rensen}, L.~M. {Duan}, J.~I. {Cirac}, and P. {Zoller}, Nature {\bf
  409},  63  (2001).

\bibitem{GioLloMac04}
V. Giovannetti, S. Lloyd, and L. Maccone, Science {\bf 306},  1330  (2004).

\bibitem{JoShiPre07}
G.~B. Jo {\it et~al.}, Phys. Rev. Lett. {\bf 98},  030407  (2007).

\bibitem{EstGroObe08}
J. Est\`{e}ve {\it et~al.}, Nature {\bf 455},  1216  (2008).

\bibitem{AppWinPol09}
J. Appel {\it et~al.}, Proc.\ Nat.\ Acad.\ Sci. {\bf 106},  10960  (2009).

\bibitem{SchLerVul0810.2582}
M.~H. Schleier-Smith, I.~D. Leroux, and V. Vuleti\'{c}, arXiv:0810.2582
  (2009).

\bibitem{ChuSchRai05}
C.~S. Chuu {\it et~al.}, Phys. Rev. Lett. {\bf 95},  260403  (2005).

\bibitem{ItaVekSte0903.3282}
A. Itah {\it et~al.}, arXiv:0903.3282  (2009).

\bibitem{LukFleZol01}
M.~D. Lukin {\it et~al.}, Phys. Rev. Lett. {\bf 87},  037901  (2001).

\bibitem{BriMolSaf07}
E. Brion, K. {M\o lmer}, and M. Saffman, Phys. Rev. Lett. {\bf 99},  260501
  (2007).

\bibitem{SafMol09}
M. Saffman and K. {M\o lmer}, Phys. Rev. Lett. {\bf 102},  240502  (2009).

\bibitem{MulLesZol09}
M. M{\"u}ller {\it et~al.}, Phys. Rev. Lett. {\bf 102},  170502  (2009).

\bibitem{HeiRaiPfa07}
R. Heidemann {\it et~al.}, Phys. Rev. Lett. {\bf 99},  163601  (2007).

\bibitem{UrbJohSaf09}
E. Urban {\it et~al.}, Nat Phys {\bf 5},  110  (2009).

\bibitem{GaeMirGra09}
A. Gaetan {\it et~al.}, Nat. Phys. {\bf 5},  115  (2009).

\bibitem{Man82}
L. Mandel, Opt. Commun. {\bf 42},  437  (1982).

\bibitem{Lou84}
R. Loudon, Opt. Commun. {\bf 49},  67  (1984).

\bibitem{Hil87}
M. Hillery, Opt. Commun. {\bf 62},  135  (1987).

\bibitem{GilKni93}
L. Gilles and P.~L. Knight, Phys.\ Rev.~A {\bf 48},  1582  (1993).

\bibitem{MoeBoeVer96}
A.~J. Moerdijk, H.~M. J.~M. Boesten, and B.~J. Verhaar, Phys.\ Rev.~A {\bf 53},
   916  (1996).

\bibitem{EsrGreBur99}
B.~D. Esry, C.~H. Greene, and J.~P. Burke, Phys. Rev. Lett. {\bf 83},  1751
  (1999).

\bibitem{NorBalBra06}
A.~A. Norrie, R.~J. Ballagh, C.~W. Gardiner, and A.~S. Bradley, Phys.\ Rev.~A
  {\bf 73},  043618  (2006).

\bibitem{BurGhrWie97}
E.~A. Burt {\it et~al.}, Phys. Rev. Lett. {\bf 79},  337  (1997).

\bibitem{SodGueDal99}
J. S\"{o}ding {\it et~al.}, Appl.\ Phys.\ B {\bf 69},  257  (1999).

\bibitem{TolOHaHuc04}
B. {Laburthe-Tolra} {\it et~al.}, Phys. Rev. Lett. {\bf 92},  190401  (2004).

\bibitem{B:vanKampen07}
N.~G. {van Kampen}, {\em Stochastic processes in physics and chemistry}, 3rd
  ed. (Elsevier, Amsterdam, 2007).

\bibitem{GerWhiSpr07}
R. Gerritsma {\it et~al.}, Phys.\ Rev.~A {\bf 76},  033408  (2007).

\bibitem{WhiGerSpr09}
S. Whitlock, R. Gerritsma, T. Fernholz, and R.~J.~C. Spreeuw, New J. Phys. {\bf
  11},  023021  (2009).

\end{thebibliography}

\end{document}